\documentclass[11pt,a4paper,final]{iopart}

\usepackage{iopams}  
\usepackage{graphicx}

\usepackage{color,subfigure}
\usepackage{bm}
\usepackage{epsfig}
\usepackage{graphicx}
\usepackage{amssymb,amsbsy,amsgen,amsfonts,amstext}
\usepackage{dcolumn}
\usepackage{mathrsfs}
\usepackage{latexsym}
\usepackage{array}

\newcommand{\braket}[2]{\langle#1|#2\rangle}
\newcommand{\bra}[1]{\left\langle{#1}\right\vert}
\newcommand{\ket}[1]{\left\vert{#1}\right\rangle}
\newcommand{\ketbra}[2]{|#1\rangle \langle#2|}

\begin{document}

\title{Quasiparticle Localisation via Frequent Measurements}

\author{David A. Herrera-Mart\'i}
\address{Centre for Quantum Technologies, National University of Singapore, 3 Science Drive 2, Singapore 117543, Singapore}
\ead{dahm@nus.edu.sg}

\author{Ying Li}
\address{Centre for Quantum Technologies, National University of Singapore, 3 Science Drive 2, Singapore 117543, Singapore}

\author{Leong Chuan Kwek$^{1,2,3}$}
\address{$^1$Centre for Quantum Technologies, National University of Singapore, 3 Science Drive 2, Singapore 117543, Singapore}
\address{$^2$Institute of Advanced Studies, Nanyang Technological University, 60 Nanyang View Singapore 639673, Singapore}
\address{$^3$National Institute of Education, 1 Nanyang Walk Singapore 637616, Singapore}

\begin{abstract}
Topological quantum memories in two dimensions are not thermally stable, since once a quasiparticle excitation is created, it will delocalise at no energy cost. This places an upper bound on the lifetime of quantum information stored in them. We address this issue for a topological subsystem code introduced in [Bravyi S.{\em et al.}, Quant.~Inf.~Comp.~13(11):0963]. By frequently measuring the gauge operators of the code, a technique known as Operator Quantum Zeno Effect [Li Y. {\em et al.}, preprint arXiv:1305.2464], the dynamics responsible for quasiparticle motion is suppressed, and can eventually be ``frozen" in the large frequency limit. A feature of this method is that the density operator of the code does not commute with the measurement operators, so the density matrix will be randomised after a few measurements. However the logical operators commute with the measurement operators and are thus protected.
\end{abstract}

\vspace{2pc}

\maketitle

\section{Introduction}

The theory of quantum error correcting codes provides the means to designing a quantum memory, which is to be regarded as the basic building block of future quantum computers. Topological quantum codes have proved to be specially advantageous in pushing progress in this enterprise, since they feature high thresholds and rely on local interactions. In order to protect quantum information from noise, two complementary strategies have been studied.

On one hand, a quantum memory should be designed to be robust against fluctuations of external parameters, which means that the information would need to be encoded in degrees of freedom which are very weakly coupled to the environment. Ideally, this decoupling from environmental fluctuations should increase as the memory is made larger, which, due to the local nature of noise, suggests that information should be stored in non-local degrees of freedom. One possibility to achieve this is to design a quantum system whose degenerate ground state consists of codewords of some error correcting code, so that local errors are energetically penalised. The first and most salient example of such a memory is the toric code \cite{Kita97b,dennis2002topological}, in which information is stored in a degenerate ground space, and logical errors are non-contractible loops of single-qubit errors. Topological order prevents the existence of thermally stable memories in two dimensions \cite{landon2013local, bravyi2009no}. Intuitively, this is because for two-dimensional topological memories coupled to a thermal bath, a local error can evolve induced by unwanted terms of the Hamiltonian and turn into non-trivial error chain at no energy cost. It is possible to design thermally stable memories in three dimensions \cite{bravyi2011energy} and even in four dimensions \cite{alicki2010thermal}. However it remains unclear how these devices would be implemented and controlled. Of particular relevance to this work are the results in where several aspects about thermal fragility and localisation of local errors in two-dimensional memories have been considered \cite{wootton2011bringing,stark2011localization, kay2011capabilities, brown2013glassy}.

On the other hand, since the Hamiltonian of a quantum memory will in general contain extra terms due to unwanted imperfections, impurities or external fields, among others, undesired evolution of the quantum information will take place.  This needs to be accounted for actively, that is, quantum error correction techniques have to be used to undo the unwanted evolution \cite{pastawski2010limitations}. In order to perform active error correction, the stabilisers of the quantum code have to be frequently measured to obtain a syndrome vector, which indicates which operations must be carried out to undo the errors.

Future quantum memories will use both active error correction as well as some amount of protection at the physical level. The more protection gained at the physical level, the less resources in terms of redundancy, measurements and classical post-processing will be needed. 

Here we explain how to supplement topological memories with ideas from a recently proposed  technique by some of the authors, known as  Operator Zeno Effect \cite{wang2013operator,li2013quantum}. We focus on two-dimensional memories, although this method is applicable to any code that features a subsystem structure \cite{bacon2006operator, poulin2005stabilizer}. Among the benefits of this new approach are easier syndrome obtention, since only three qubit measurements are needed, as well as the suppression of the unwanted parts of the Hamiltonian resulting in logical errors.

\section{A Subsystem Toric Code} 

\begin{figure}[!h]
\centering
\begin{subfigure}[]
\centering
\includegraphics[width=0.7\textwidth]{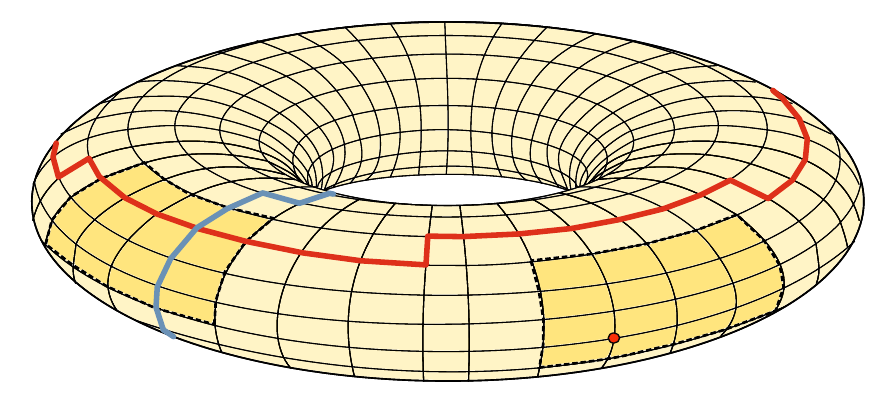}
\end{subfigure}\hspace{0cm}\\
\begin{subfigure}[]
\centering
\includegraphics[width=0.35\textwidth]{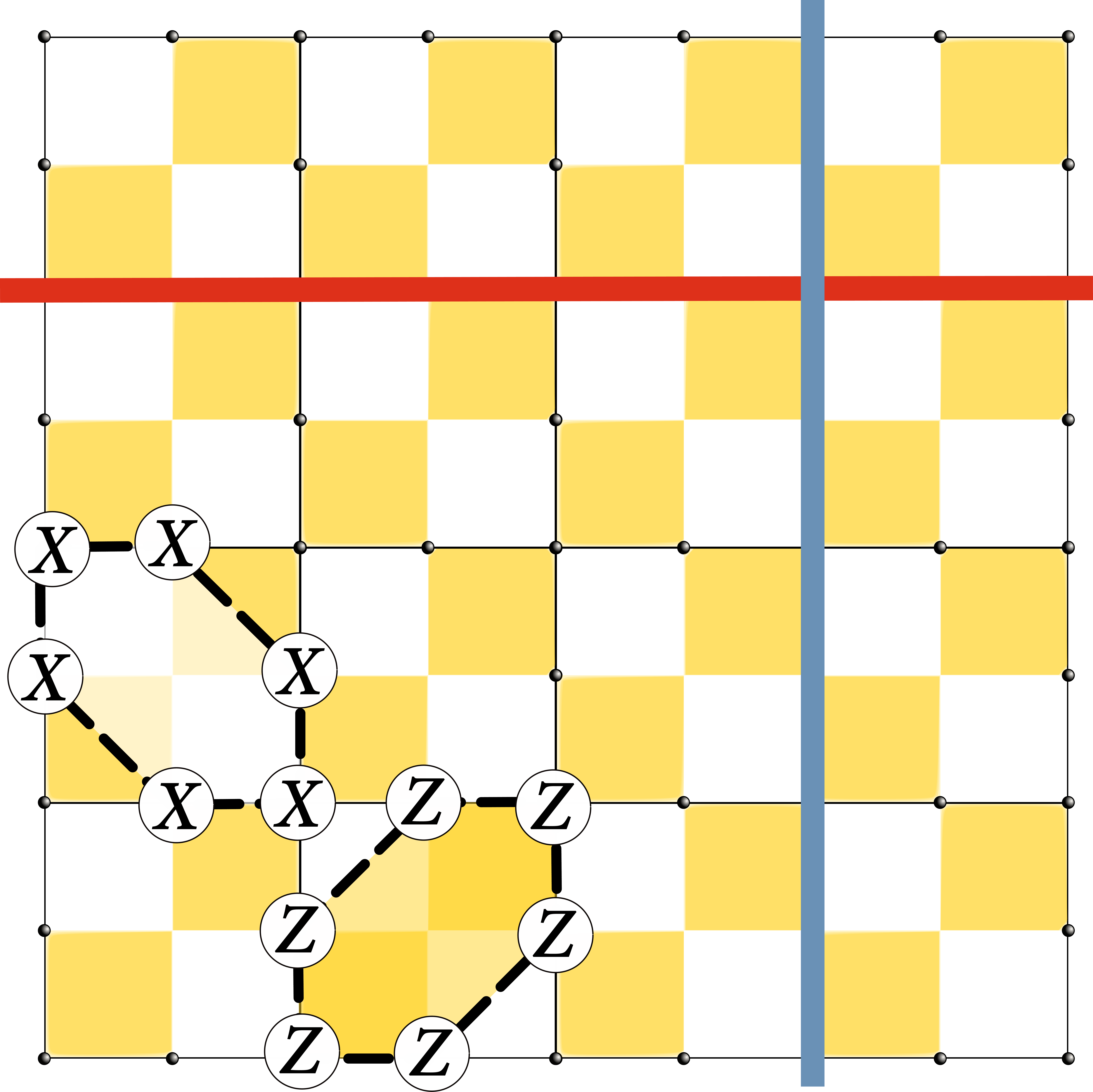}
\end{subfigure}\hspace{2cm}
\begin{subfigure}[]
\centering
\includegraphics[width=0.35\textwidth]{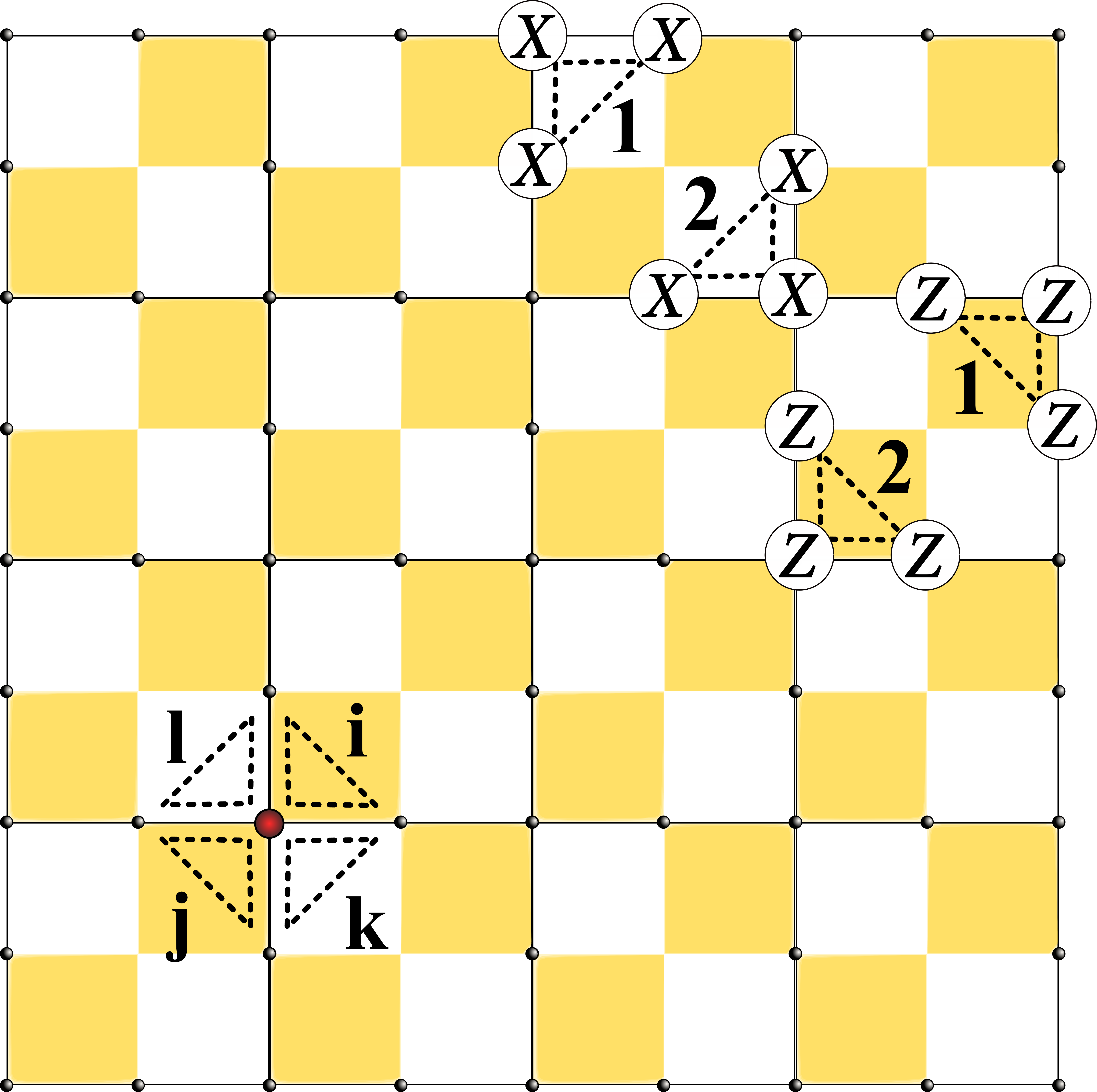}
\end{subfigure}
\caption{\label{Code}{\bf(a)} Periodic boundary conditions allow to identify logical operators with non-contractible loops of single qubit operators. The loops blue and red correspond to two such operators. {\bf(b)} [Zoom on left] The virtual subspace is stabilised by the operators defined in the dark ($Q^Z_i$) and light ($Q^X_i$) plaquettes. These plaquettes will share an even number of physical qubits, so they commute. Anticommutation of the logical operators is ensured since the logical loops always share an odd number of physical qubits. {\bf(c)} [Zoom on right] Each plaquette operator is decomposed into two gauge operators acting on three physical qubits, $Q^Z_k = g^Z_{k,1} g^Z_{k,2}$, and $Q^X_k = g^X_{k,1} g^X_{k,2}$.  Anticommutation of gauge operators belonging to overlapping plaquettes determine the algebra of a local gauge qubit. The red dot indicates the existence of an operator that would induce tunnelling of a quasiparticle between the plaquettes. Frequently measuring the gauge operators that act on that qubit will hinder this transition (see Eq.~(\ref{Tunnel2})).}
\end{figure}

The original toric code Hamiltonian \cite{Kita97b} has a degenerate ground state which is used to encode two logical qubits. The simplest excitations of the code correspond to the creation of a pair of quasiparticles. Once they are created, they can move within the code at no energy cost, and recombination of these quasiparticles after a non-contractible orbit results in a non-trivial action on the encoded qubits. 
To see the relation between the usual toric code and the \emph{subsystem} toric code, consider the merging of the original toric code plaquette stabilisers as shown in Fig.~\ref{Code}(b). This operation removes one qubit and two stabiliser constraints, so that the subspace stabilised by the new plaquettes is now larger. However, for each removed qubit, introducing gauge operators will induce a local virtual subsystem structure that preserves the topological protection of the code. This construction is equivalent to the one presented in \cite{bravyi2012subsystem}.

The Hamiltonian of the new code is\\

\begin{equation}
H_\textrm{STC} = -\frac{\Delta}{2}\big(\sum Q^Z_i+ \sum Q^X_j\big)  
\label{Hamiltonian}
\end{equation}
where the new plaquettes are $Q^Z_i = \bigotimes_{\kappa\in Q^\textrm{dark}_i }Z_\kappa$ and $Q^X_j = \bigotimes_{\kappa\in Q^\textrm{light}_j }X_\kappa$ are six-body terms. Greek indices are used for qubit sites, whereas latin indices are used for plaquettes.

The new code consists of $n=3L^2$ physical qubits and $m = L^2 - 2g$ independent plaquettes, where $g$ is the genus of the underlying manifold (for the torus, $g=1$).  The common eigenspace of all the plaquette stabilisers is called the virtual subspace ${\mathcal V}$ and it has dimension $\textrm{dim}({\mathcal V}) = 2^{n-m}$. The subsystem structure arises from the identification of the operators $\tilde X_k$'s and $\tilde Z_k$'s  associated to each removed qubit  $k$ as gauge generators \cite{zanardi2004quantum, poulin2005stabilizer}. Note that these gauge generators are not single-qubit operators, but rather weight-three operators. The $2L^2$ gauge generators, together with the $4g$ non-trivial orbits, induce a partition of the virtual subspace, $\mathcal{V} = \bigotimes_k \mathcal{G}_k\bigotimes_l \mathcal{L}_l$, where $\{k\}$ is a string of length $L^2$ and $\{l\}$ is a string of length $2g$. For all logical operators $\bar L \in \mathcal{L}_l$, $[\tilde X_k,\bar L] = [\tilde Z_k,\bar L] = 0\textrm{  }\forall k$. In fact, $\bar L \equiv \tilde X_k\bar L \equiv \tilde Z_k\bar L\textrm{  }\forall k,l$. Crucially, the gauge qubits associated with the removed physical qubits are localised, whereas the logical subsystems retain its non-local character. 

A common feature of subsystem codes is that gauge operators can be used to ``break up" the stabiliser operators, facilitating thus the measurement process \cite{bombin2010topological}. Specifically, for the new plaquettes centred around gauge qubit $k$,

\begin{eqnarray}
Q^Z_k &=& (Q^Z_k \tilde Z_k) \tilde Z_k = Z^{\otimes 3} Z^{\otimes 3}\equiv g^Z_{k,1} g^Z_{k,2},\\
Q^X_k &=& (Q^X_k \tilde X_k) \tilde X_k = X^{\otimes 3} X^{\otimes 3} \equiv g^X_{k,1} g^X_{k,2},
\end{eqnarray}
where the gauge operators $g^Z_{k,l}$, $g^X_{k,l}$ are defined via the gauge qubits, $\tilde Z_k$ and $\tilde X_k$. Note that for each plaquette, only one of the gauge operators is linearly independent. Only three-qubit measurements, rather than six-qubit measurements are needed in order to retrieve a syndrome vector. It has been shown in \cite{bravyi2012subsystem} that this code features a threshold of almost a percent against depolarising errors.

\section{Suppression of Quasiparticle Motion}

We now show how, by frequently measuring the gauge operators of the subsystem toric code, random motion of excitations can be slowed down and even ``frozen", so that the unwanted dynamics leading to logical errors will not be present in the effective Hamiltonian of the code.

\subsection{Operator Quantum Zeno Effect}

The Operator Quantum Zeno Effect (OQZE) is a refinement of the Subspace Quantum Zeno Effect \cite{li2013quantum}, where a set of non-commuting observables are measured sequentially, in such a way that the quantum state evolves randomly, instead of being constrained within one subspace of the total Hilbert space. The evolution superoperator for a time $\tau$ is:\\
\begin{equation}
\centering
\mathcal{U}_N(\tau) = [\mathcal{U}(\frac{\tau}{N})\mathcal{P}]^N,
\label{Evolution}
\end{equation}
where $\varrho(\tau/N)=\mathcal{U}(\tau/N)\varrho(0) = e^{-i\frac{H\tau}{N}} \varrho(0) e^{i\frac{H\tau}{N}}$. $H$ is the system Hamiltonian and the superoperator sequence $\mathcal{P} = \mathcal{P}^{(1)}\mathcal{P}^{(2)}\dots\mathcal{P}^{(K)}$ is composed of non-selective measurement superoperators $\mathcal{P}^{(j)}\varrho = \sum_q M^{(j)}_q\varrho M^{(j)\dagger}_q$, although measurements can also be selective (see subsection~\ref{Selective}). In general, $M^{(i)}_q$ and $M^{(j)}_{q'}$ do not commute, and this constitutes the generalisation of the OQZE. Yet, certain operators fulfilling the condition $\mathcal{P}A = A$, which we call Measurement Invariant Operators (MIO),  will slow down their evolution as the frequency of the measurement series increases. Interestingly, a MIO can still evolve driven by a Hamiltonian which is itself a MIO. This can be seen as an effective evolution arising as the frequency of the measurements is increased:\\

\begin{equation}
\varrho(\tau)=\mathcal{U}_N(\tau) \varrho(0) \stackrel{N\rightarrow\infty}{\longrightarrow}  e^{-i\frac{\tilde H\tau}{N}} \varrho(0) e^{i\frac{\tilde H\tau}{N}},
\label{Dynamics}
\end{equation}
where  $\tilde H = \lim_{N\rightarrow\infty} \frac{1}{N}\sum^N_m \mathcal{P}^m H$ is the effective Hamiltonian resulting from frequently performing the measurements \cite{li2013quantum}. 
Figuratively speaking, the frequent measurements will ``sever" the code Hamiltonian until commutes with all the measurement operators, as it were. 

In general, the advantage of the OQZE over the Subspace QZE is that it is possible to freeze dynamics outside of the computational subspace by measuring operators acting on fewer qubits. This is a vital consideration when it comes to physical implementations of a quantum memory, since reducing the physical requirements will revert in lower effective error rates.

\subsection{Single-Exciton Subspaces}

An effective dynamics will emerge as a result of frequently measuring the gauge operators, as summarised in Eq.~(\ref{Dynamics}). The key idea is to separate the total Hamiltonian into measurement invariant and measurement non-invariant parts, $ H_T = H_\textrm{STC} + H_\textrm{MNI}$. The measurement invariant part is composed of the subsystem toric code stabilisers, which commute with the gauge operators, and the measurement non-invariant part $H_\textrm{MNI} = \delta \sum A_\kappa$, with $A_\kappa = \alpha_\kappa X_\kappa + \beta_\kappa Y_\kappa + \gamma_\kappa Z_\kappa$ accounts for local imperfections and/or external fields. $\alpha_\kappa,\beta_\kappa,\gamma_\kappa$ are random variables such that $\alpha^2_\kappa+\beta^2_\kappa+\gamma^2_\kappa= 1$, so in general the operators $A_\kappa$ will not commute with the stabiliser operators. Therefore they can induce transitions between states with different energies, that is, they can create quasiparticle excitations. However these transitions will be suppressed if $\Delta \gg \delta$. We henceforth set $\delta = 1$ for simplicity.

Transitions between same energy states are not suppressed by the gap, so once a quasiparticle is created, it will move around the surface of the code at no energy cost. If untreated, this motion will result in logical errors after a time which increases at least logarithmically with the code distance \cite{kay2011capabilities}. Although excitations come in pairs, we will be concerned with motion of a single quasiparticle. To do that, we look at the single-exciton subspace of $H_\textrm{STC}$. In the subsystem toric code, a plaquette operator is read indirectly through measuring its associated gauge operators, which entails a degeneracy in the definition of states with one excitation. In a  code of dimensions $L\times L$ a given configuration in the single-exciton subspace can be written

\begin{equation}
\ket{Q^Z_k, {\bf m}} = \ket{0,m_1}_1\otimes\ket{0, m_2}_2\otimes \cdots \ket{1,m_k}_k \cdots \otimes\ket{0, m_{L^2}}_{L^2},
\label{Microstates}
\end{equation}
where ${\bf m} = [m_1, m_2, \dots m_{L^2}]$ is a chain of measurement outcomes of the gauge operators. Specifically
\begin{eqnarray}
\ket{0,m}_i &=& \ket{m}_{i,1}\ket{m}_{i,2},\\
\ket{1,m}_i &=& \ket{m}_{i,1}\ket{m\oplus1}_{i,2}.
\end{eqnarray}

Here $(-1)^mg^Z_{i,l}\ket{m}_{i,l} = \ket{m}_{i,l}$, with $l \in \{1,2\}$, $m \in \{0,1\}$. $m$ denotes the measurement outcome of gauge operator $(i,l)$ and  ``$\oplus$" is addition modulo 2 (see Fig.~\ref{Code}).  The same applies to light plaquettes.

Each state corresponding to a quasiparticle at one location can be written as many equivalent configurations. We call these configurations ``information microstates". The degeneracy in the choice of the state with one excitation at one site is $2^{L^2}$, and it is a consequence of the subsystem structure of the code. In a similar way one can define the ground state in the subdomain, using a slightly modified version of Eq.~(\ref{Microstates}). Once the code boundaries are considered, the topological degeneracy is recovered. This entails a macroscopic degeneracy of the topologically protected ground states (see Summary and Discussion).\\

The single-exciton subspaces are associated to the projectors $\Pi_Z = \sum_{k, {\bf m}} \ketbra{Q^Z_k, {\bf m}}{Q^Z_k, {\bf m}}$ and $\Pi_X = \sum_{k, {\bf m}}\ketbra{Q^X_k, {\bf m}}{Q^X_k, {\bf m}}$. The dimension of each of these subspaces is $\textrm{rank }\Pi_Z=\textrm{rank }\Pi_X=L^2\times2^{L^2}$. Notice that $\Pi_Z$ and $\Pi_X$ do not commute, due to the subsystem structure of the code. However, the two types of quasiparticle excitations can be treated independently, so the single-exciton Hamiltonian ca be written as


\begin{eqnarray}\label{Truncation}
H_{1e} = \Pi_Z H_T\Pi_Z + \Pi_X H_T\Pi_X  \\
= \sum_{i,{\bf m}}  \mu^Z_{i{\bf m}}\ket{Q^Z_i, {\bf m}}\bra{Q^Z_i, {\bf m}}  + \sum_{i,{\bf m}}  \mu^X_{i{\bf m}}\ket{Q^X_i, {\bf m}}\bra{Q^X_i, {\bf m}} + & \nonumber\\
+ \sum_{i, {\bf m}, \atop j\in N(i), {\bf m'}}  t^Z_{i {\bf m}, j {\bf m'}}\ket{Q^Z_i, {\bf m}}\bra{Q^Z_j, {\bf m}'} + \sum_{i, {\bf m}, \atop j\in N(i), {\bf m'}} t^X_{i {\bf m}, j {\bf m'}}\ket{Q^X_i, {\bf m}}\bra{Q^X_j, {\bf m}'}, &\nonumber
\end{eqnarray}
where onsite energies are $\mu^Z_{i{\bf m}} = \bra{Q^Z_i, {\bf m}} H_\textrm{MNI} \ket{Q^Z_i, {\bf m}}=\sum_{\kappa\in Q^\textrm{dark}_i } \bra{Q^Z_i, {\bf m}} \gamma_\kappa Z_\kappa \ket{Q^Z_i, {\bf m}}$, $\mu^X_{i{\bf m}} = \bra{Q^X_i, {\bf m}} H_\textrm{MNI} \ket{Q^Z_i, {\bf m}}=\sum_{\kappa\in Q^\textrm{light}_i } \bra{Q^X_i, {\bf m}} \alpha_\kappa X_\kappa \ket{Q^X_i, {\bf m}}$, and the diagonal terms associated to $H_\textrm{STC}$ have been obviated. Since the measurement outcomes of {\bf m} are random, $\mu^Z_{i{\bf m}} =\sum_{\kappa\in Q^\textrm{dark}_i }(-1)^{m_i} \gamma_\kappa$, and  $\mu^X_{i{\bf m}} =\sum_{\kappa\in Q^\textrm{light}_i }(-1)^{m_i} \alpha_\kappa$.  Remarkably, the subsystem structure of the code will induce onsite disorder in the single-exciton subspace, and this disorder will moreover fluctuate in time. This is a consequence of the non-commutativity of the gauge operators.

The tunneling elements in Eq.~(\ref{Truncation}) are $t^Z_{i {\bf m}, j {\bf m'}}= \bra{Q^Z_i, {\bf m}}H_\textrm{MNI}\ket{Q^Z_j, {\bf m}'}$ and $t^X_{i {\bf m}, j {\bf m'}}= \bra{Q^X_i, {\bf m}}H_\textrm{MNI}\ket{Q^X_j, {\bf m}'}$.  Similarly to the onsite energies, the tunnelling amplitudes will depend on the measurement outcomes:

\begin{eqnarray}
t^Z_{i {\bf m}, j {\bf m'}}=  \sum_\kappa\bra{Q^Z_i, {\bf m}}A_\kappa \ket{Q^Z_j, {\bf m}'} = \bra{Q^Z_i, {\bf m}}A_{\kappa_{ij}}\ket{Q^Z_j, {\bf m}'}&  \\
= \bra{0,m_1}_1\cdots \bra{1,m_i}_i \cdots \bra{0,m_{L^2}}_{L^2}A_{\kappa_{ij}} \ket{0,m'_1}_1\cdots \ket{1,m'_j}_j \cdots \ket{0,m'_{L^2}}_{L^2}& \nonumber \\
= \bra{m_i}_{i,1}\bra{m_i\oplus1}_{i,2}\bra{m_j}_{j,1}\bra{m_j}_{j,2}A_{\kappa_{ij}}(\tau)\ket{m'_i}_{i,1}\ket{m'_i}_{i,2} \ket{m'_j}_{j,1}\ket{m'_j\oplus1}_{j,2}&\nonumber\\
=(\bra{m_i\oplus1}_{i,2}\bra{m_j}_{j,1}A_{\kappa_{ij}}\ket{m'_i}_{i,2} \ket{m'_j}_{j,1})(\braket{m_i}{m'_i}_{i,1}\braket{m_j}{m'_j\oplus1}_{j,2})&\nonumber \\
= (\alpha_{\kappa_{ij}} + i(-1)^{m'_i+m'_j}\beta_{\kappa_{ij}})\delta_{m_i,m'_i}\delta_{m_j,m'_j\oplus1} \nonumber
\end{eqnarray}

where $\kappa_{ij}$ denotes the location of qubit shared by plaquettes  $i$ and $j$. In the third line, indices of irrelevant plaquettes have been contracted. In the fourth line we observe that $A_{\kappa_{ij}}$ only acts on subsystems $(i,2)$ and $(j,1)$ (see Fig.~\ref{Code}). Without loss of generality, we discuss only tunnelling between dark plaquettes, since the same reasoning applies to light plaquettes. 

This randomness in the tunnelling elements, together with the randomness in the onsite energy, will lead to some amount of localisation \cite{kramer1993localization}. However, in next subsection we show that the dynamics caused by the measurement non-invariant Hamiltonian is slowed-down are even cancelled in the large measurement frequency limit, so both the tunneling elements and the onsite disorder caused by $H_\textrm{MNI}$ will vanish As the measurement frequency is increased.


%

\subsection{Localisation via Frequent Measurements}

The time evolution is given by the superoperator in Eq.~(\ref{Evolution}), so the transition amplitude between two states with the same energy after a time $\tau$, in the Heisenberg picture, is given by 

\begin{eqnarray}\label{Tunnel1}
t^Z_{i {\bf m}, j {\bf m'}} (\tau) = \bra{Q^Z_i, {\bf m}}A_{\kappa_{ij}}(\tau)\ket{Q^Z_j, {\bf m}'}&  \\
= \bra{0,m_1}_1\cdots \bra{1,m_i}_i \cdots \bra{0,m_{L^2}}_{L^2}A_{\kappa_{ij}} (\tau)\ket{0,m'_1}_1\cdots \ket{1,m'_j}_j \cdots \ket{0,m'_{L^2}}_{L^2}& \nonumber \\
= \bra{m_i}_{i,1}\bra{m_i\oplus1}_{i,2}\bra{m_j}_{j,1}\bra{m_j}_{j,2}A_{\kappa_{ij}}(\tau)\ket{m'_i}_{i,1}\ket{m'_i}_{i,2} \ket{m'_j}_{j,1}\ket{m'_j\oplus1}_{j,2}&\nonumber
\end{eqnarray}
%


For each operator $A_{\kappa_{ij}}(\tau)=[\mathcal{P}^\dag\mathcal{U}(\frac{-\tau}{N})]^NA_{\kappa_{ij}}$, the measurements of only four adjacent gauge operators must be considered (see Fig.~\ref{Code}(c)). Thus, the series of non-selective measurements  is written $\mathcal{P} = \dots\mathcal{P}^{(l, 2)}_X\mathcal{P}^{(j, 1)}_Z\mathcal{P}^{(k, 1)}_X\mathcal{P}^{(i, 2)}_Z\dots$; namely, each qubit site is measured four times. The measurement superoperators are $\mathcal{P}^{(j, 1)}_Z \varrho = \sum_{q\in\{0,1\}} M^{(j, 1)}_q\varrho M^{(j, 1)\dag}_q$, with $M^{(j, 1)}_0 = \sqrt{\lambda}\frac{ I + g^Z_{j,1}}{2}+\sqrt{1-\lambda}\frac{ I - g^Z_{j,1}}{2}$ and  $M^{(j, 2)}_1 = \sqrt{\lambda}\frac{ I - g^Z_{j,1}}{2}+\sqrt{1-\lambda}\frac{ I + g^Z_{j,1}}{2}$ such that the completeness relation $\sum_{q\in\{0,1\}}  M^{(j, 1)\dag}_q M^{(j, 1)}_q =  I$ holds. Equivalently, $\mathcal{P}^{(k,1)}_X \varrho = \sum_{q\in\{+,-\}} M^{(k,1)}_q\varrho M^{(k,1)\dag}_q$, with $M^{(k,1)}_+ = \sqrt{\lambda}\frac{ I + g^X_{k,1}}{2}+\sqrt{1-\lambda}\frac{ I - g^X_{k,1}}{2}$, $M^{(k,1)}_- = \sqrt{\lambda}\frac{ I - g^X_{k,1}}{2}+\sqrt{1-\lambda}\frac{ I + g^X_{k,1}}{2}$. Here the parameter $\lambda$ determines the \emph{weakness} of the measurement. 

The tunnelling amplitude is a monotonic decreasing function of the measurement frequency. To see this explicitly, consider the last line of Eq.~(\ref{Tunnel1}) and relabel $\ket{q^Z_i}=\ket{m_i}_{i,1}\ket{m_i\oplus1}_{i,2} \ket{m_j}_{j,1}\ket{m_j}_{j,2}$ and $\ket{q^Z_j}=\ket{m'_i}_{i,1}\ket{m'_i}_{i,2} \ket{m'_j}_{j,1}\ket{m'_j\oplus1}_{j,2}$. The absolute value of the tunnelling amplitude can then be rewritten as

\begin{eqnarray}\label{Tunnel2}
|t^Z_{i {\bf m}, j {\bf m'}} (\tau)|=  |\braket{q^Z_i}{A_{\kappa_{ij}}(\tau)|q^Z_j}| \leq ||q^Z_i||_2||A_{\kappa_{ij}}(\tau)q^Z_j||_2 \leq ||q^Z_i||_2||A_{\kappa_{ij}}(\tau)||&&\nonumber \\
 \leq ||q^Z_i||_2||A_{\kappa_{ij}}(\tau)||_1 = ||\mathcal{P}^\dag \mathcal{U}(\frac{-\tau}{N})\cdots\mathcal{P}^\dag \mathcal{U}(\frac{-\tau}{N})A_{\kappa_{ij}}||_1&&\nonumber \\
 \leq ||\mathcal{U}(\frac{-\tau}{N})\cdots\mathcal{P}^\dag \mathcal{U}(\frac{-\tau}{N})A_{\kappa_{ij}}||_1&&\nonumber \\
 \vdots \nonumber\\
 \leq ||\mathcal{P}^\dag \mathcal{U}(\frac{-\tau}{N})A_{\kappa_{ij}}||_1&&\nonumber \\
 \leq ||\mathcal{U}(\frac{-\tau}{N})A_{\kappa_{ij}}||_1&&\nonumber \\
 = ||A_{\kappa_{ij}}||_1&&
\end{eqnarray}

where in the first line we have resorted to the definition of the operator norm $||\bullet||$. In the second line, monotonicity between the operator norm and the trace norm $||\bullet||_1$ has been used. Since we had traced out the indices of the irrelevant plaquettes, using the trace norm to bound the strength of $A_{\kappa_{ij}}(\tau)$ is justified. Each subsequent inequality has been obtained by considering the action of measurement superoperators on Hermitian operators. Note that any Hermitian operator $\Lambda$ can be decomposed as $\Lambda=\Lambda_+ - \Lambda_-$, where $\Lambda_\pm$ are positive operators corresponding to the positive and negative eigenvalues of $\Lambda$, respectively. Then, $||\mathcal{P}\Lambda ||_1 \leq ||\mathcal{P}\Lambda_+||_1 + ||\mathcal{P}\Lambda_-||_1 = \textrm{Tr}(\mathcal{P}\Lambda_+) + \textrm{Tr}(\mathcal{P}\Lambda_-) = \textrm{Tr}(\mathcal{P}(\Lambda_+ + \Lambda_-))= \textrm{Tr}(\Lambda_+ - \Lambda_-) \leq ||\Lambda||_1$.

\begin{figure}[!t]
\centering
\includegraphics[width=14.0 cm]{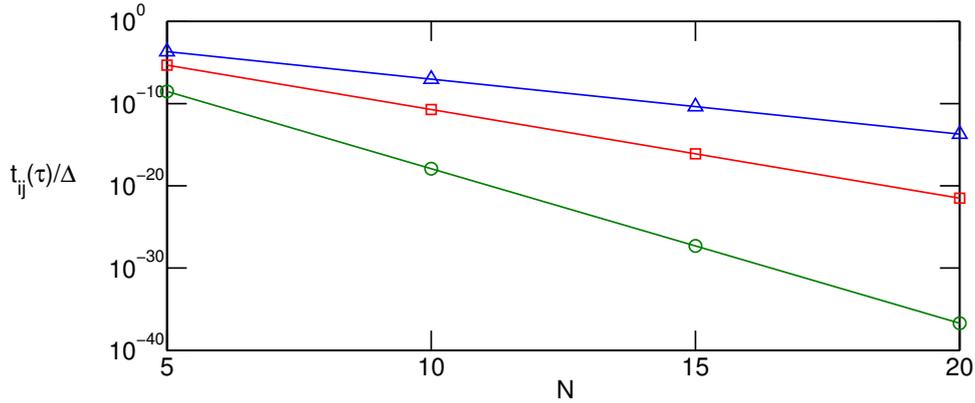}
\caption{\label{Tunnelling}After a time $\tau$, the tunnelling amplitude $t_{ij}(\tau)$ decreases as function of the measurement frequency $N/\tau$. Our numerics show an exponential suppression of the tunnelling amplitude of the exponential using weak projective measurements. The different lines denote the ``weakness" of the mesasurement: blue (triangles) stands for $\sqrt{\lambda} = 0.9$, red (squares) is $\sqrt{\lambda} = 0.95$ and green (circles) is for $\sqrt{\lambda} = 0.99$. Weak measurements constitute a model for faulty read-out procedures, where a wrong outcome in the syndrome measurement is induced by a random error. The bounds derived in Eq.~\ref{Tunnel2} hold for more general measurements. }
\end{figure}

The last inequalities in Eq.~(\ref{Tunnel2}) are tight iff $[M^{(k,l)}_q,\cdots\mathcal{P}^\dag \mathcal{U}(\frac{-\tau}{N})A_{\kappa_{ij}}] = 0$ $\forall q,k,l$, {\it i.e.} all the measurements are $M^{(k,l)}_q = \frac{ I}{2}$ $\forall q,k,l$. The numerical results shown in Fig.~\ref{Tunnelling} indicate the dependence on the ``weakness" of the measurements. This applies to every tunnelling element and onsite energy in the code. Furthermore, the same reasoning can be used to argue that the onsite disorder vanishes as the measurement frequency increases.

Frequently performing non-selective measurements of the gauge operators will hinder transitions between same energy states. Due to the non-commutativity of the measurement operators, after each time interval $\tau$ the density matrix in the single-exciton subspace will be driven into a mixture of ``information microstates". This entails that coherent tunnelling between same energy eigenstates will not be possible, so the motion of the quasiparticle will be diffusive. Logical operators, on the other hand, commute with all the measurement operators, and will be conserved. Note that this is at stark contrast with what would happen in a code without a subsystem structure, since then the density matrix would live in the logical subspace and therefore would commute with all measurement operators.

\subsection{Selective Measurements}\label{Selective}

A syndrome vector has to be obtained periodically, which implies that rounds of selective measurements interspersed by the non-selective measurements have to be performed with a given frequency. The frequency of syndrome extraction will be inversely proportional to the non-selective measurement frequency increases, since the excitations will move at lower speed. It is not difficult to see from the decomposition $\Lambda=\Lambda_+ - \Lambda_-$ that selective measurements cannot increase the norm of operators and thus will also satisfy the inequalities in Eq.~(\ref{Tunnel2}).


\section{Distortion-Induced Localisation}

In previous section we introduced weak projective measurements as a model for errors occurring at read-out stage. It is possible to consider yet another deviation from the idealised measurement scenario by allowing the measured gauge operators to be slightly different from the ideal ones. This can be the consequence of some source of systematic errors, or can arise as a deliberate strategy to further protect the information from random motion of the excitations, as we now show.

We particularise for dark plaquettes without loss of generality, since the same reasoning applies to light plaquettes. A distorted gauge $g'^{Z}_{k,l}$ operator can be written as

\begin{equation}
g'^{Z}_{k,l} = \sqrt{1-\epsilon}g^{Z}_{k,l} + \sqrt{\epsilon}e^Z_{k,l},
\end{equation}
where $\epsilon\geq 0$ is a small random variable, and $e^Z_{k,l}$ is a unitary operator such that $\{g^{Z}_{k,l},e^Z_{k,l}\} = 0$. Equivalently, $g^{Z}_{k,l} = \sqrt{1-\epsilon}g'^{Z}_{k,l} + \sqrt{\epsilon}e'^Z_{k,l}$, where $e'^Z_{k,l} = \sqrt{\epsilon}g^{Z}_{k,l}  - \sqrt{1-\epsilon}e^Z_{k,l}$. The constraints on the distorted gauge operators are

\begin{eqnarray}
\{g'^{Z}_{i,l},g'^{X}_{i,l}\} &=& 0 \hspace{5mm}\forall i,l \\
\lbrack g'^{Z}_{i,l},g'^{Z}_{j,l}\rbrack &=& 0 \hspace{5mm}\forall i,j,l \\
\lbrack g'^{X}_{i,l},g'^{X}_{j,l}\rbrack &=& 0\hspace{5mm}\forall i,j,l \\
\lbrack g'^{Z}_{i,l},g'^{X}_{j,l}\rbrack &=& 0 \hspace{5mm}\forall i\neq j,l 
\end{eqnarray}

These conditions ensure that the subsystem structure is maintained even in the presence of distortion.  For each plaquette, the distortion introduced by the new measurements is quantified by  $d^Z_{k,l}(\epsilon) = ||g^{Z}_{k,l}-g'^{Z}_{k,l}|| = 1-\sqrt{1-\epsilon}\leq\sqrt{\epsilon}$. To see the effect of distortion in the high measurement frequency limit, the diagonal elements due to $H_\textrm{MNI}$ in Eq.~(\ref{Truncation}) cannot be dismissed anymore. Assuming a measurement frequency high enough such that the effect of $H_\textrm{MNI}$ can be neglected, we focus on the expectation value of the plaquette operators, which now are site-dependent:


\begin{eqnarray}
 \mu^Z_{i{\bf m}} (\tau)= \frac{-\Delta}{2} \bra{ Q^Z_i, {\bf m}} Q^Z_i(\tau)\ket{Q^Z_i, {\bf m}} \nonumber\\
\approx \frac{-\Delta}{2} \bra{Q^Z_i, {\bf m}} \mathcal{P'}^NQ^Z_i\ket{Q^Z_i, {\bf m}} = \frac{-\Delta}{2} \bra{Q^Z_i, {\bf m}}  \mathcal{P'}^Ng^{Z}_{i,1}g^{Z}_{i,2}\ket{Q^Z_i, {\bf m}}\nonumber\\
= -\frac{\Delta\sqrt{1-\epsilon}}{2} \bra{Q^Z_i, {\bf m}} g'^{Z}_{i,1}g^{Z}_{i,2}\ket{Q^Z_i, {\bf m}}  -\frac{\Delta\sqrt{\epsilon}}{2}\bra{Q^Z_i, {\bf m}} \mathcal{P'}^Ne'^{Z}_{i,1}g^{Z}_{i,2}\ket{Q^Z_i, {\bf m}}\nonumber\\
=-\frac{\Delta(1-\epsilon)}{2} \bra{Q^Z_i, {\bf m}} g^{Z}_{i,1}g^{Z}_{i,2}\ket{Q^Z_i, {\bf m}} -\frac{\Delta\sqrt{\epsilon}}{2}\bra{Q^Z_i, {\bf m}} \mathcal{P'}^Ne'^{Z}_{i,1}g^{Z}_{i,2}\ket{Q^Z_i, {\bf m}}\nonumber\\
\stackrel{N\rightarrow\infty}{\longrightarrow} -\frac{\Delta(1-\epsilon)}{2}  \bra{Q^Z_i, {\bf m}} Q^Z_i\ket{Q^Z_i, {\bf m}} =  -\frac{\Delta(1-\epsilon)}{2} \hspace{5mm} \forall {\bf m},
\label{Onsite}
\end{eqnarray}
where in the second line we have considered that $\Delta\gg\delta$ so the plaquette is never flipped by $H_\textrm{MNI}$. In the third line follows from $\mathcal{P'}g'^{Z}_{i,1} = g'^{Z}_{i,1}$ and in fourth line we observe that $e^Z_{i,1}$ is non-diagonal in the single-exciton basis. Finally, in order to obtain the limit, we have used Eq.~(\ref{Dynamics}), proved in \cite{li2013quantum}, and the fact that $\{g'^{Z}_{i,1},e'^{Z}_{i,1}\} = 0$. Similar arguments hold for  $\mu^X_i (\tau)$.

\begin{figure}[!h]
\centering
\begin{subfigure}[]
\centering
\includegraphics[width=1\textwidth]{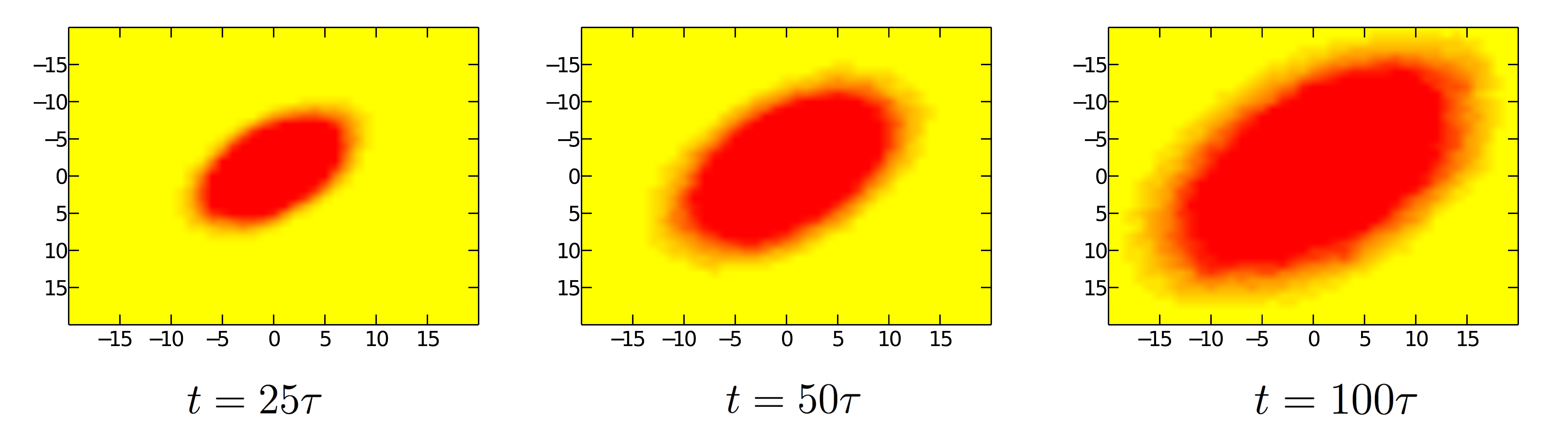}
\end{subfigure}\hspace{0cm}\\
\begin{subfigure}[]
\centering
\includegraphics[width=1\textwidth]{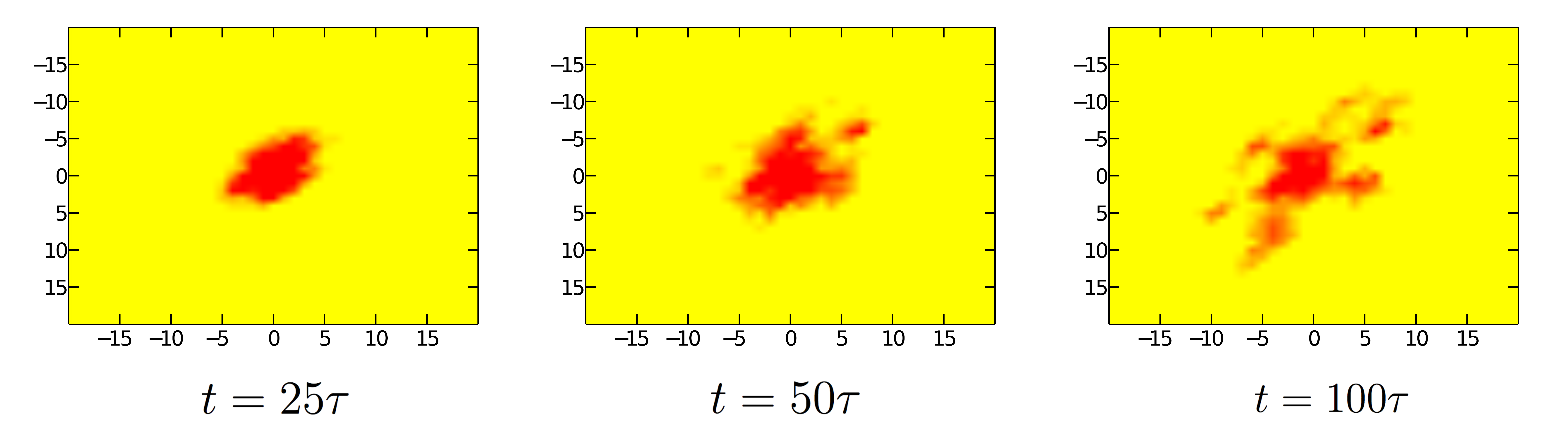}
\end{subfigure}
\caption{\label{Localisation}The motion of a quasiparticle excitation in two different scenarios, averaged over 100 simulations. The simulations were carried out for a $50\times50$ code, for dark plaquettes.  {\bf(a)} Scenario with no disorder. The measurements are perfect and the quasiparticle undergoes evolution dictated by $H_\textrm{T}$. {\bf(b)} Distorted measurements lead to a disordered scenario with an site-dependent energy. In the simulations we set the variation of onsite energies to be $\sigma_{\mu^Z_i} = 10t^Z_{ij}$.}
\end{figure}

As one example of a measurement strategy that can induce disorder and localisation in the high frequency limit, consider the action of $R_X(\theta) = e^{-i\frac{X\theta}{2}}$ on a single qubit of the gauge operator $g^Z_{(i,l)}$, such that $g'^Z_{(i,l)} = R_X(\theta) g^Z_{(i,l)}R^\dagger_X(\theta) = \cos\theta g^Z_{(i,l)} - \sin\theta (Z\otimes Y\otimes Z)$. Similarly, one can consider rotations in different basis.  If all plaquettes acting on the qubit undergo the same rotation, then the conditions on the distorted gauge operators are fulfilled. A random choice of angles for each qubit site will lead to a random onsite potential, as shown in Eq.~(\ref{Onsite}). It has been argued that any amount of disorder will lead to localisation \cite{kramer1993localization, wootton2011bringing, stark2011localization, kay2011capabilities}. Numerical simulation of quasiparticle motion confirmed that distorted measurements will lead to localisation in the high frequency limit, as illustrated in Fig.~\ref{Localisation}.

\section{Summary and Discussion}

We showed that by frequently measuring the gauge operators of a subsystem toric code, evolution of quasiparticle excitations can be slowed down, and even detained if distortion is introduced.  One advantage over the {\em subspace} toric code is that in this case only three-qubit measurements are needed for syndrome extraction. This is relevant for reducing the effective error probability at read-out stage.


{\em Validity at non-zero temperature}--- Our discussion has been centred around {\em coherent motion} of quasiparticles due to imperfections of the code, which is to be the main source of errors at zero temperature. 
At non-zero temperature a  bath will induce stochastic motion that will lead to quasiparticle diffusion, in which the non-unitary tunnelling rates can be modelled using Lindblad-type terms. We would like to remark that the technique here presented can {\em in principle} be used to slow down non-unitary dynamics, provided the measurement frequency is larger than the bath's typical frequency.  Furthermore, the disorder introduced by distorted measurements will persist in the large measurement frequency limit and will lead to localisation at non-zero temperature.

{\em Geometrical frustration and subsystem codes}---It is useful to imagine error correction as a process whereby entropy is extracted from the code by reducing some fictitious temperature, so that after a successful round of error correction, the logical qubit is left in a pure state with vey high probability \cite{steane2002quantum}. This is in contrast to what happens in the subsystem toric code: even after a large number of successful rounds of error correction, the code will retain some amount of {\em residual entropy}. This residual entropy is a result of the choice in the definition of the ground states and as a consequence, it increases exponentially in the system's size.

The same phenomenon arises in frustrated systems \cite{moessner2006geometrical}. In this context, the impossibility of satisfying all Hamiltonian terms at once is analogous to the existence of non-commuting gauge operators. Although the Hamiltonian in Eq.~(\ref{Hamiltonian}) is frustration-free, frequently measuring the gauge operators will drive the density matrix into a mixture of ``information microstates". This exciting connection between certain subsystem codes as and some frustrated systems is made clear if thermal entropy is substituted by information entropy.

Incidentaly, although the code introduced in Eq.~(\ref{Hamiltonian}) had been already described in \cite{bravyi2012subsystem}, we arrived to the conclusions presented in this work starting from a different point. Specifically, we considered two copies of a two-dimensional projection of the pyrochlore lattice \cite{runge2004charge,fulde2012correlated,castelnovo2008magnetic} as an inspiration for the subsystem toric code, where non-frustrated loops in the lattice are mapped to our stabilisers. It is known that the pyrochlore exhibits a macroscopic degeneracy of its ground state at zero temperature \cite{moessner2006geometrical}, and this can be regarded as the thermal counterpart to the logical degeneracy of the subsystem toric code.


{\em Acknowledgments}---D.H.M., Y.L. and L.C.K. acknowledge support from the National Research Foundation \& Ministry of Education, Singapore. We thank D. Poulin for valuable comments and pointing out errors on an earlier version of the manuscript. Also, we thank B. Brown and T. Rudolph for critical reading of the manuscript.
\section*{References}
\bibliographystyle{unsrt}

\end{document}